\documentclass[12pt]{article}
\usepackage{a4wide,epsfig,psfrag,amsmath,amssymb,cite,scalefnt}
\usepackage{color}

\graphicspath{ {figs/} }

\newcommand{\gsim}{\;\rlap{\lower 3.5 pt \hbox{$\mathchar \sim$}} \raise 1pt
 \hbox {$>$}\;}
\newcommand{\lsim}{\;\rlap{\lower 3.5 pt \hbox{$\mathchar \sim$}} \raise 1pt
 \hbox {$<$}\;}

\allowdisplaybreaks

\begin{document}

\title{\vskip-3cm{\baselineskip14pt
    \begin{flushleft}
      \normalsize SFB/CPP-14-19\\
      \normalsize TTP14-009 \\
      \normalsize DESY 14-038
  \end{flushleft}}
  \vskip1.5cm
  Hadronic contribution to the muon anomalous magnetic moment to 
  next-to-next-to-leading order
}

\author{
  Alexander Kurz$^{a,b}$,
  Tao Liu$^{a}$,
  Peter Marquard$^{b}$,
  Matthias Steinhauser$^{a}$
  \\[1em]
  {\small\it (a) Institut f{\"u}r Theoretische Teilchenphysik,}
  {\small\it Karlsruhe Institute of Technology (KIT)}\\
  {\small\it 76128 Karlsruhe, Germany}
  \\
  {\small\it (b) Deutsches Elektronen Synchrotron (DESY),}\\
  {\small\it 15738 Zeuthen, Germany}
}

\date{}

\maketitle

\thispagestyle{empty}

\begin{abstract}

  We compute the next-to-next-to-leading order hadronic 
  contribution to the
  muon anomalous magnetic moment originating from 
  the photon vacuum polarization. The corresponding three-loop kernel
  functions are calculated using asymptotic expansion techniques 
  which lead
  to analytic expressions. Our final result,  
  $a_\mu^{\rm had,NNLO} = 1.24 \pm 0.01 \times 10^{-10}$,
  has the same order of
  magnitude as the current uncertainty of the leading order hadronic
  contribution and should thus be included in future analyses.

  \medskip

\end{abstract}

\thispagestyle{empty}


\newpage


\section{Introduction}

The anomalous magnetic moment of the electron and the muon
are measured with high precision and at the same time
also accurately predicted including high-order 
quantum corrections (see, e.g., 
Refs.~\cite{Melnikov:2006sr,Jegerlehner:2008zza,Jegerlehner:2009ry,Miller:2012opa}
for reviews on this topic).
Notable recent achievements in this context are the 
five-loop QED corrections obtained in Refs.~\cite{Aoyama:2012wj,Aoyama:2012wk}.

In the case of the muon the largest input to the uncertainty comes from
hadronic contributions which to a large extent rely on experimental
measurements of the cross section $\sigma(e^+e^-\to\mbox{hadrons})$.  Several
groups have performed the leading order
(LO)~\cite{Davier:2010nc,Hagiwara:2011af,Jegerlehner:2011ti,Benayoun:2012wc}
and next-to-leading order
(NLO)~\cite{Krause:1996rf,Greynat:2012ww,Hagiwara:2003da,Hagiwara:2011af}
analysis. In this paper we compute the next-to-next-to-leading order (NNLO)
hadronic corrections to the anomalous magnetic moment of the electron and the 
muon. We evaluate
the three-loop kernels in the limit $M_\mu \ll m_\pi$ and show that
four expansion terms are sufficient to obtain a precision
far below the per cent level.  Note that we do not consider the light-by-light
contribution where the external photon couples to the hadronic loop (see,
e.g., Ref.~\cite{Prades:2009tw}) but only the contributions involving the
hadronic vacuum polarizations.

In the next Section we briefly mention some technical details of
our calculation and discuss the NLO contribution.
Section~\ref{sec::NNLOmu} contains the results of the various 
NNLO contributions for the muon anomalous magnetic moment
and in Section~\ref{sec::NNLOe} we apply our results
to the anomalous magnetic moment of the electron.
We conclude in Section~\ref{sec::concl}.


\section{Technicalities and NLO result}

\begin{figure}[t]
  \begin{center}
  \begin{tabular}{cccc}
    \includegraphics[scale=0.65]{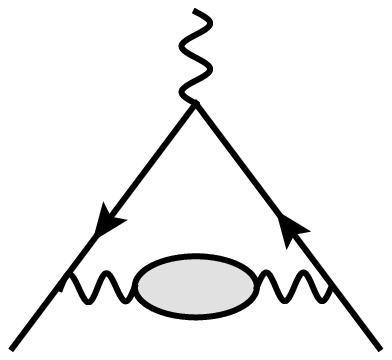} &
    \includegraphics[scale=0.65]{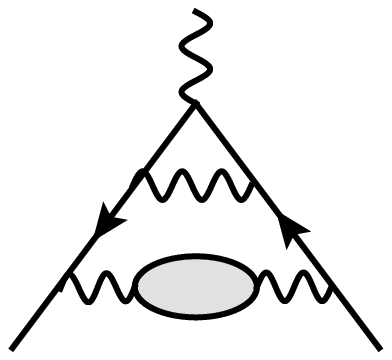} &
    \includegraphics[scale=0.65]{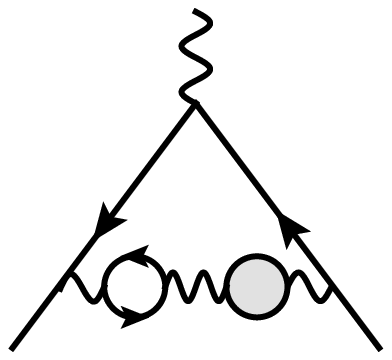} &
    \includegraphics[scale=0.65]{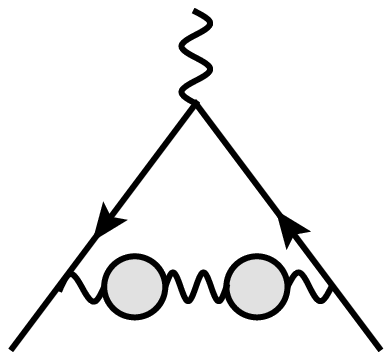}
    \\ (a) LO & (b) 2a & (c) 2b & (d) 2c \\
  \end{tabular}
  \caption{\label{fig::FD1}LO and sample NLO Feynman diagrams contributing to
    $a_\mu^{\rm had}$.}
  \end{center}
\end{figure}

The LO hadronic contribution to the anomalous magnetic moment of the muon
(see Fig.~\ref{fig::FD1}) can be computed via
\begin{eqnarray}
  a_\mu^{(1)} &=& \frac{1}{3} \left(\frac{\alpha}{\pi}\right)^2
  \int_{m_\pi^2}^\infty {\rm d} s \frac{R(s)}{s} K^{(1)}(s)
  \,,
  \label{eq::aLO}
\end{eqnarray}
where $\alpha$ is the fine structure constant and
$R(s)$ is given by the properly normalized total 
hadronic cross section in electron positron collisions
\begin{eqnarray}
  R(s) &=& \frac{ \sigma(e^+e^-\to\mbox{hadrons}) }{ \sigma_{pt} }
  \,,
\end{eqnarray}
with $\sigma_{pt} = 4\pi\alpha^2/(3s)$.
A convenient integral representation for $K^{(1)}(s)$ is given by
\begin{eqnarray}
  K^{(1)}(s) &=& 
  \int_0^1 {\rm d} x \frac{x^2(1-x)}{x^2 + (1-x) \frac{s}{M_\mu^2}}
  \,,
\end{eqnarray}
analytic results can be found in Refs.~\cite{BroRaf68,Eidelman:1995ny}.

A crucial input for the evaluation of $a_\mu^{\rm had}$ is a compilation of
the experimental data for $R(s)$ as obtained by various experiments.  In our
analysis we use a {\tt FORTRAN} code which is provided to us by the authors of
Ref.~\cite{Hagiwara:2011af}. This gives us access to both the central values
and the upper and lower limit of $R(s)$.  However, the use of the latter leads
to a vast overestimation of the final uncertainty since we have no information
about the correlations of the individual data points.  Thus, we use a
heuristic method and consider the difference between $a_\mu^{\rm had}$ as
obtained from the central and upper or lower limit of $R(s)$ and divide it by
three which leads to realistic (and still conservative) error estimates at LO
and NLO.  In fact, for the energy region $[0.32~\mbox{GeV},1.43~\mbox{GeV}]$
we obtain the LO contribution $608.19 \pm 3.97 \times 10^{-10}$ which is in
a good agreement with $606.50 \pm 3.35 \times 10^{-10}$ from table~5 of
Ref.~\cite{Hagiwara:2011af}.  Note that in this paper we do not aim for an
improved prediction of the LO or NLO contribution. Rather we present for the
first time NNLO hadronic predictions. Obviously, for that purpose, the
described prescription for the determination of the uncertainty is
sufficient.

The contribution to $a_\mu$ from the $J/\Psi$, $\Psi(2S)$ and $\Upsilon(nS)$
($n=1,\ldots,4$) resonances is obtained with the help of the
narrow-width approximation as described in Ref.~\cite{Hagiwara:2003da}.

At NLO three different contributions are distinguished as shown in
Fig.~\ref{fig::FD1}(b), (c) and (d).  We have computed the 
kernels $K^{(2a)}$ and $K^{(2b)}$
using the methods of asymptotic expansion~\cite{Smirnov:2013} and in
that way confirmed the results provided in Ref.~\cite{Krause:1996rf}. 
Ref.~\cite{Krause:1996rf} also contains analytic expressions for
$K^{(2c)}(s,s^\prime)$. It is, however, convenient to work with the
one-dimensional integral representation which reads~\cite{Krause:1996rf}
\begin{eqnarray}
  K^{(2c)}(s,s^\prime) &=& 
  \int_0^1 {\rm d} x \frac{x^4(1-x)}
  {\left[x^2 + (1-x) \frac{s}{M_\mu^2}\right]
    \left[x^2 + (1-x) \frac{s^\prime}{M_\mu^2}\right]}
  \,.
\end{eqnarray}

The contributions $a_\mu^{(2a)}$ and $a_\mu^{(2b)}$ are obtained from
Eq.~(\ref{eq::aLO}) after replacing $K^{(1)}$ by either $K^{(2a)}$ or
$K^{(2b)}$ and $(\alpha/\pi)^2$ by $(\alpha/\pi)^3$.
$a_\mu^{(2c)}$ requires an integration over both $s$ and $s^\prime$
and is obtained from
\begin{eqnarray}
  a_\mu^{(2c)} &=& \frac{1}{9} \left(\frac{\alpha}{\pi}\right)^3
  \int_{m_\pi^2}^\infty {\rm d} s {\rm d} s^\prime 
  \frac{R(s)}{s} \frac{R(s^\prime)}{s^\prime} K^{(2c)}(s,s^\prime)
  \label{eq::cNLO}
  \,.
\end{eqnarray}

Our results for the three contributions read
\begin{eqnarray}
  a_\mu^{(2a)} &=& -20.90 \times 10^{-10}\,, \nonumber\\
  a_\mu^{(2b)} &=&  10.68 \times 10^{-10}\,, \nonumber\\
  a_\mu^{(2c)} &=&   0.35 \times 10^{-10}\,,
\end{eqnarray}
which leads to
\begin{eqnarray}
  a_\mu^{\rm had,NLO} &=& -9.87 \pm 0.09 \times 10^{-10}\,,
\end{eqnarray}
in a good agreement with Refs.~\cite{Hagiwara:2003da,Hagiwara:2011af}.


\section{\label{sec::NNLOmu}NNLO hadronic contributions to $a_\mu$}

\begin{figure}[t]
  \begin{center}
  \begin{tabular}{cccc}
    \includegraphics[scale=0.65]{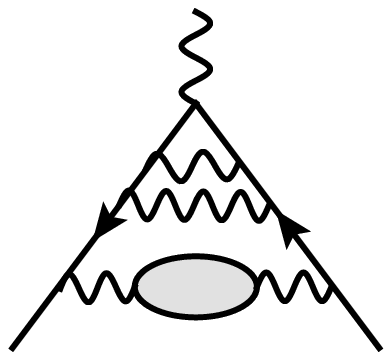} &
    \includegraphics[scale=0.65]{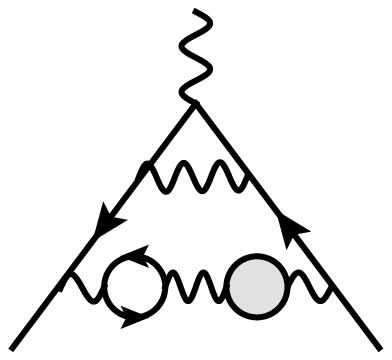} &
    \includegraphics[scale=0.65]{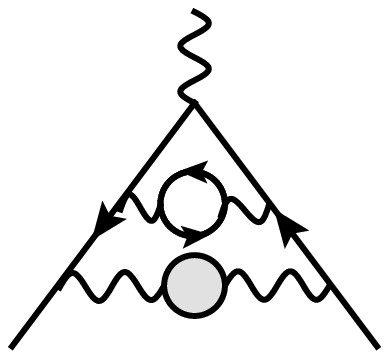} &
    \includegraphics[scale=0.65]{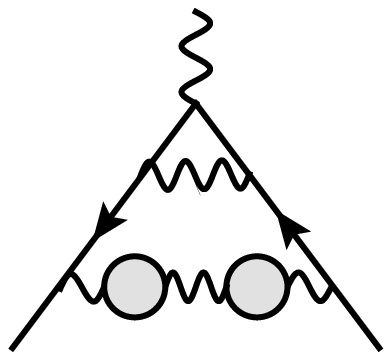}
    \\ (a) $3a$ & (b) $3b$ & (c) $3b$ & (d) $3c$ \\
    \includegraphics[scale=0.65]{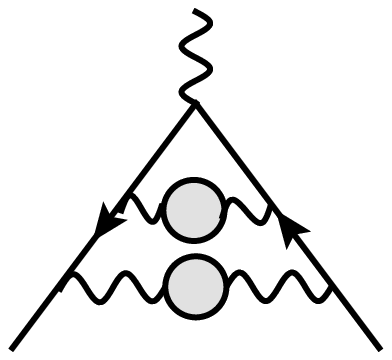} &
    \includegraphics[scale=0.65]{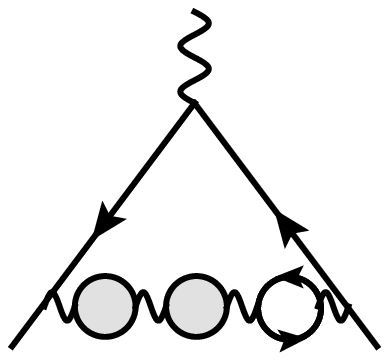} &
    \includegraphics[scale=0.65]{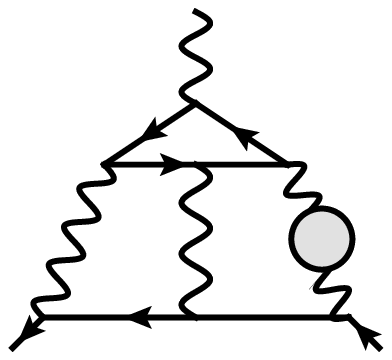} &
    \includegraphics[scale=0.65]{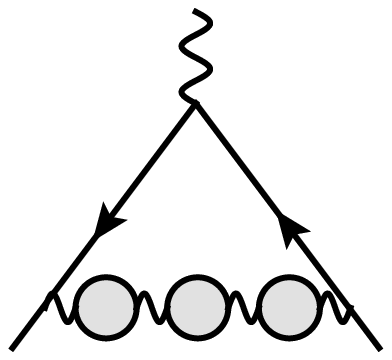}
    \\ (e) $3c$ & (f) $3c$ & (g) $3b$,lbl & (h) $3d$ \\
  \end{tabular}
  \caption{\label{fig::FD_nnlo}Sample NNLO Feynman diagrams contributing to
    $a_\mu^{\rm had}$. The external fermions are muons and the 
      fermions in the closed loops represent electrons.}
  \end{center}
\end{figure}

We classify the NNLO contributions in analogy to 
NLO according to the number of hadronic insertions and
closed electron loops. This leads to five different kernels
which contain the following contributions
(see Fig.~\ref{fig::FD_nnlo} for sample Feynman diagrams):
\begin{itemize}
\item $K^{(3a)}$: one hadronic insertion;
  up to two additional photons to the LO Feynman
  diagram; contains also the contributions with one or
    two closed muon loops and the light-by-light-type diagram
    with a closed muon loop.
\item $K^{(3b)}$: one hadronic insertion and one or two
  closed electron loops and additional photonic corrections;
  the external photon couples to the muon.
\item $K^{(3b,\rm lbl)}$:
  light-by-light-type contribution with
  closed electron loop and one hadronic insertion;
  the external photon couples to the electron.
\item $K^{(3c)}$: two hadronic insertions and additional
  photonic corrections and/or closed electron or muon loops.
\item $K^{(3d)}$: three hadronic insertions.
\end{itemize}
Note that we do not consider contributions with closed
tau lepton loops since they are suppressed by an additional factor
$M_\mu^2/M_\tau^2$. Actually, at NLO these contributions
amount to $0.01\times 10^{-10}$ and thus
we anticipate that the corresponding NNLO terms are
even smaller.

The calculation of $K^{(3a)}(s)$ proceeds in analogy to the corresponding one-
and two-loop cases: we apply an asymptotic expansion for $\sqrt{s}\gg M_\mu$
and compute terms up to order $(M_\mu^2/s)^4$.
The
minimal value of $\sqrt{s}$ is given by $m_\pi$ and thus the 
largest value of the expansion
parameter is $M_\mu^2/m_\pi^2 \approx 0.6$. Note, however, that
the contribution from the energy interval $[m_\pi,2m_\pi]$ is very small
such that in practice the expansion parameter is 
$M_\mu^2/(4m_\pi^2) \approx 0.15$  or smaller for higher energies.
We observe a good
convergence of the series as can be seen by considering the difference for
$a_\mu^{(3a)}$ ($a_\mu^{(3b)}$) computed from $K^{(3a)}(s)$ ($K^{(3b)}(s)$) by
including and neglecting the highest available term which is at the per mil
level.  For $K^{(3b)}$ and $K^{(3b,\rm lbl)}$ we consider in addition the
limit $M_\mu \gg M_e$ and compute terms up to quartic order in
$M_e$. Corrections of order $M_e/M_\mu$ or higher turn out to be
negligibly small.  In the case of $K^{(3b)}$ the leading term  for
$M_e\to0$ can be obtained using renormalization group techniques (see,
e.g., Ref.~\cite{Lee:2013sx} where four-loop correction to $a_\mu$ with closed
electron loops have been considered). However, a non-zero electron mass is
crucial for the light-by-light-type contribution $K^{(3b,\rm lbl)}$ since the
Feynman integrals are divergent in case $M_e=0$ is chosen. Thus, a non-trivial
asymptotic expansion has to be applied.  The latter is realized with the help
of the program {\tt asy}~\cite{Pak:2010pt,Jantzen:2012mw}.

\begin{figure}[tb]
  \begin{center}
    \begin{tabular}{cc}
      \includegraphics[scale=0.3]{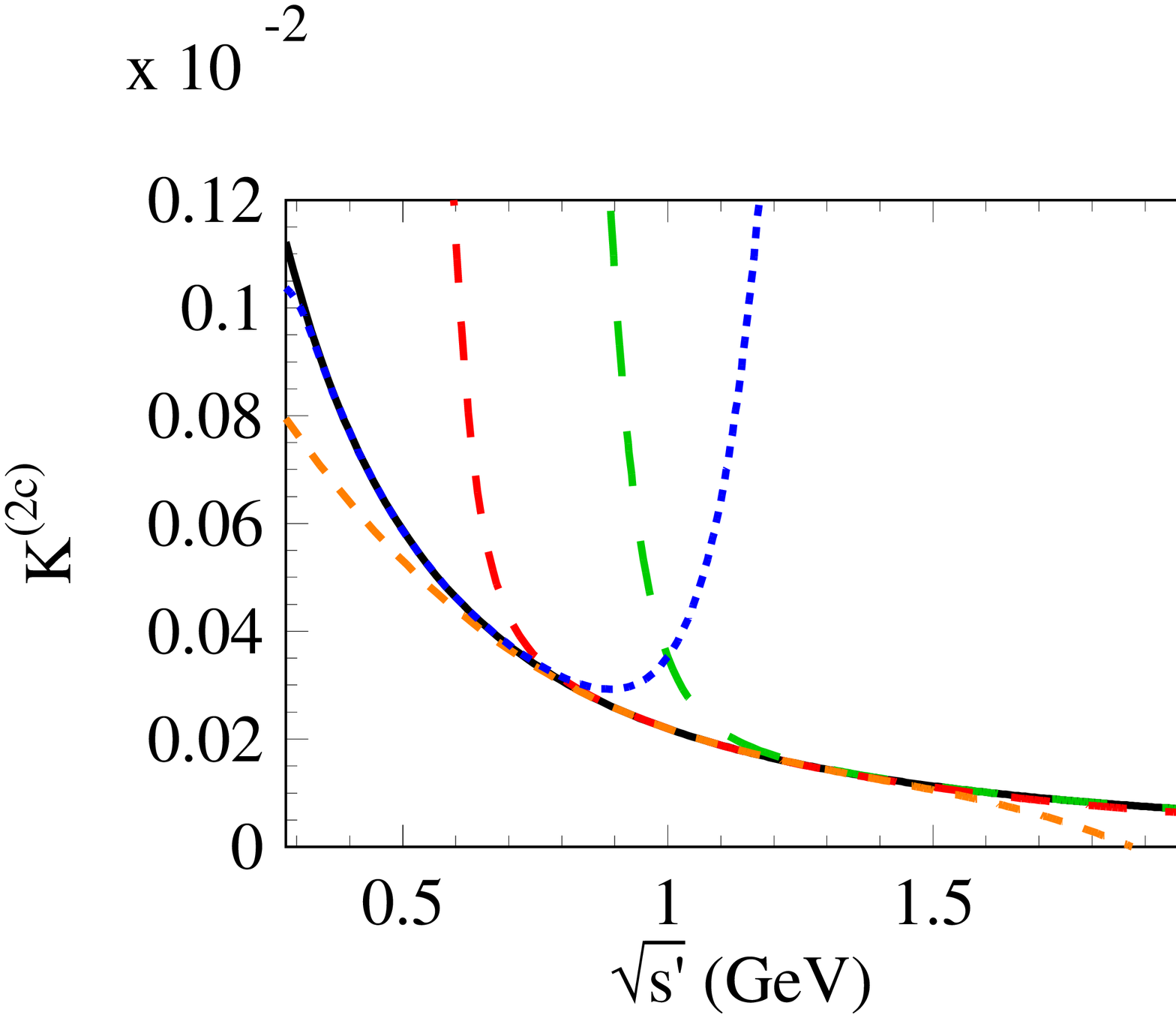} &
      \includegraphics[scale=0.3]{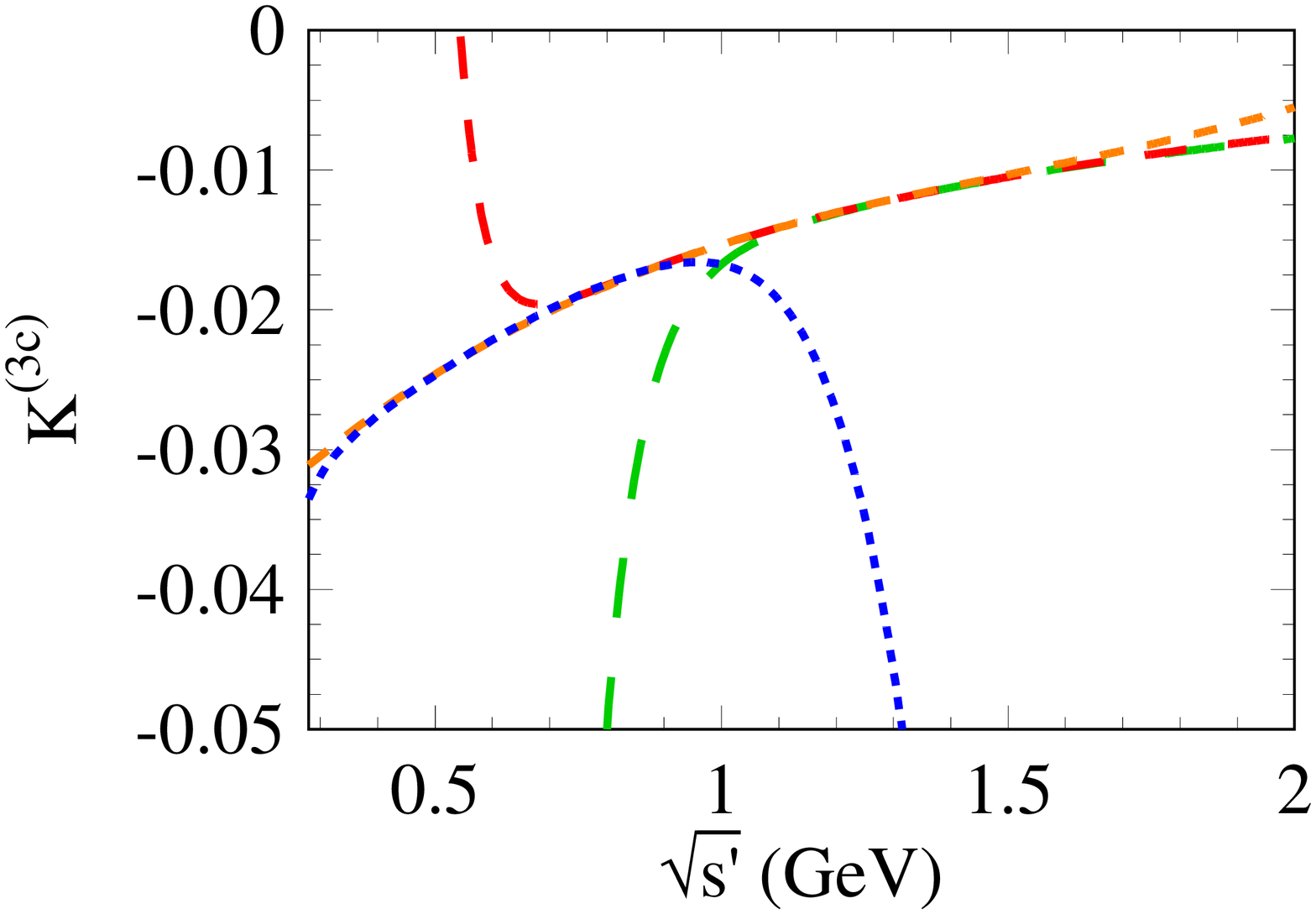}
      \\
      (a) & (b)
    \end{tabular}
    \caption{\label{fig::k2ck3c}(a) Comparison of exact result (solid, black) for
      $K^{(2c)}(s,s^\prime)$ and the various approximations for 
      $s\gg s^\prime$ (blue, dotted),
      $s\approx s^\prime$ (orange and red, short and medium dashed)
      and $s\ll s^\prime$ (green long dashed) for $\sqrt{s}=1$~GeV as a
      function of $\sqrt{s^\prime\,}$.  (b) Approximations for
      $K^{(3c)}(s,s^\prime)$.}
  \end{center}
\end{figure}

For the computation of $K^{(3c)}(s,s^\prime)$ we use asymptotic expansions in
the limits $s\gg s^\prime \gg M_\mu^2$, $s \approx s^\prime\gg M_\mu^2$ and
$s^\prime\gg s\gg M_\mu^2$ and construct an interpolating function by
combining the results from the individual limits.  This procedure can be
tested in the case of $K^{(2c)}(s,s^\prime)$ where a comparison to the exact
result is possible. In Fig.~\ref{fig::k2ck3c}(a) we show
$K^{(2c)}(s,s^\prime)$ for $\sqrt{s}=1$~GeV as a function of
$\sqrt{s^\prime\,}$.\footnote{Note that there are two curves for the
  region $s \approx s^\prime$ which correspond to the expansion parameters
  $1-\sqrt{s}/\sqrt{s^\prime}$ and $1-\sqrt{s^\prime}/\sqrt{s}$, see also
  Refs.~\cite{Eiras:2006xm,Pak:2012xr}.} (For larger values of $\sqrt{s}$ the
convergence properties are even better.)  One observes that for each value of
$\sqrt{s^\prime\,}$ there is perfect agreement between the exact result (solid
line) and at least one of the approximations (dotted and dashed lines).
Furthermore, the 
final results for $a_\mu^{(2c)}$ computed from the exact and approximated
kernels differ by less than $1\%$.

Fig.~\ref{fig::k2ck3c}(b) shows the corresponding results for
$K^{(3c)}(s,s^\prime)$. 
For each value of $s^\prime$ we have at least two approximations which agree
with each other. Thus it is evident that a function can be defined 
which agrees piecewise with one of the approximations.

For the kernel of the triple-hadronic insertion,
$K^{(3d)}(s,s^\prime,s^{\prime\prime})$, we derive
a one-dimensional integral representation
which is given by
\begin{eqnarray}
  K^{(3d)}(s,s^\prime,s^{\prime\prime}) &=& 
  \int_0^1 {\rm d} x \frac{x^6(1-x)}
  {\left[x^2 + (1-x) \frac{s}{M_\mu^2}\right]
    \left[x^2 + (1-x) \frac{s^\prime}{M_\mu^2}\right]
    \left[x^2 + (1-x) \frac{s^{\prime\prime}}{M_\mu^2}\right]}
  \,.
\end{eqnarray}

We refrain from listing explicit results for 
the NNLO kernels but provide the results in
computer-readable form on the web page~\cite{progdata}.

For the computation of $a_\mu^{(3a)}$, $a_\mu^{(3b)}$ and
$a_\mu^{(3b,\rm lbl)}$ one inserts the corresponding
kernel in Eq.~(\ref{eq::aLO}) and replaces
$(\alpha/\pi)^2$ by $(\alpha/\pi)^4$.
Furthermore, $a_\mu^{(3c)}$ is obtained from Eq.~(\ref{eq::cNLO})
with $K^{(2c)}$ replaced by $K^{(3c)}$ and $(\alpha/\pi)^3$ 
by $(\alpha/\pi)^4$ and the three-fold hadronic insertion
is calculated from
\begin{eqnarray}
  a_\mu^{(3d)} &=& \frac{1}{27} \left(\frac{\alpha}{\pi}\right)^4
  \int_{m_\pi^2}^\infty {\rm d} s {\rm d} s^\prime  {\rm d} s^{\prime\prime}
  \frac{R(s)}{s} \frac{R(s^\prime)}{s^\prime}
  \frac{R(s^{\prime\prime})}{s^{\prime\prime}} K^{(3d)}(s,s^\prime,s^{\prime\prime}) 
  \label{eq::dNNLO}
  \,.
\end{eqnarray}

For the individual NNLO contributions we obtain the 
results
\begin{eqnarray}
  a_\mu^{(3a)} &=&         0.80   \times 10^{-10} \,,\nonumber\\
  a_\mu^{(3b)} &=&        -0.41   \times 10^{-10} \,,\nonumber\\
  a_\mu^{(3b,\rm lbl)} &=& 0.91   \times 10^{-10} \,,\nonumber\\
  a_\mu^{(3c)} &=&        -0.06   \times 10^{-10} \,,\nonumber\\
  a_\mu^{(3d)} &=&         0.0005 \times 10^{-10} \,,
\end{eqnarray}
which leads to
\begin{eqnarray}
  a_\mu^{\rm had,NNLO} &=& 1.24 \pm 0.01 \times 10^{-10}\,.
  \label{eq::amuNNLO}
\end{eqnarray}
Our result is of the same order of magnitude as the
uncertainty of the LO hadronic contribution. For example,
in Ref.~\cite{Hagiwara:2011af} an uncertainty of $3.72\times 10^{-10}$
is quoted due to the statistical and systematic errors of 
the experimental data. 
Furthermore, $a_\mu^{\rm had,NNLO}$ in Eq.~(\ref{eq::amuNNLO}) is also 
of the same order of magnitude as the experimental uncertainty 
anticipated for future experiments measuring $a_\mu$ (see, e.g.,
Ref.~\cite{Venanzoni:2012qa}). 
Thus, the NNLO hadronic
corrections should be included in the comparison with
the experimental result for $a_\mu$.


\section{\label{sec::NNLOe}NNLO hadronic contributions to $a_e$}

In this Section we apply our results to the electron anomalous 
magnetic moment,
$a_e$. At LO and for $K^{(2a)}$ this means that the lepton mass has to be
interpreted as $M_e$. $K^{(2b)}$ is absent
and we have checked that $K^{(2c)}$ gives a negligible contribution (see also 
Ref.~\cite{Krause:1996rf}). 
The situation is analogous at NNLO where we only remain with $K^{(3a)}$.

At LO and NLO our results for $a_e$ read
$a_e^{\rm had, LO}  = a_e^{(1)}  = 1.877 \times 10^{-12}$ and
$a_e^{\rm had, NLO} = a_e^{(2a)} =-0.2246 \times 10^{-12}$
which is consistent with the recent analysis of
Ref.~\cite{Nomura:2012sb} where the values 
$a_e^{\rm had, LO}  = 1.866 \pm 0.011  \times 10^{-12}$ and
$a_e^{\rm had, NLO} =-0.2234\pm 0.0014  \times 10^{-12}$
have been obtained.
At NNLO we get the result\footnote{We neglect the contribution from
  $K^{(3c)}$ since it is about a factor 100 smaller than the one
  from $K^{(3a)}$. Similarly heavy-lepton contributions
  proportional to $M_e^2/M_\mu^2$ are not taken into account.}
\begin{eqnarray}
  a_e^{\rm had, NNLO} \,\,=\,\,
  a_e^{(3a)} &=& 0.028 \pm 0.001 \times 10^{-12} \,,
  \label{eq::aeNNLO}
\end{eqnarray}
which is almost three times larger than the uncertainty of $a_e^{\rm had,
    LO}$ quoted in
Ref.~\cite{Nomura:2012sb}. It is furthermore of the same order of magnitude as
the hadronic light-by-light contribution which amounts to $a_e^{\rm had,lbl} =
0.035 \pm 0.010 \times 10^{-12}$~\cite{Aoyama:2012wj}.  Note that
currently both the uncertainty in the theory prediction for $a_e$ and the
difference between theory and experiment is of order $1\times
10^{-12}$~\cite{Aoyama:2012wj} which is about a factor 40 larger than the
result given in Eq.~(\ref{eq::aeNNLO}).


\section{\label{sec::concl}Conclusions}

We have computed the NNLO hadronic vacuum polarization
corrections to the 
anomalous magnetic moment of the muon. Five different
contributions can be distinguished which are discussed individually. 
The numerically largest contribution comes from the light-by-light-type
diagram with a closed electron loop followed by the photonic
corrections and the contribution containing a closed electron two-point
function. Multiple hadronic insertions only lead to numerical
results which are much smaller.
The main result of this paper is given in Eq.~(\ref{eq::amuNNLO}).

In Ref.~\cite{Aoyama:2012wk} the theory prediction
$a_\mu^{\rm th} = 116\, 591\, 840(59) 10^{-11}$
has been compared to the experimental
result~\cite{Bennett:2006fi,Roberts:2010cj} 
$a_\mu^{\rm exp} = 116\, 592\, 089(63) 10^{-11}$
which leads to a deviation of $2.9\sigma$.
After adding our result in Eq.~(\ref{eq::amuNNLO}) 
to $a_\mu^{\rm th}$ this reduces to $2.7\sigma$.

As a by-product we have also evaluated the NNLO hadronic corrections to
$a_e$. Our result is larger than the uncertainty at LO and of the same order
as the hadronic light-by-light contribution.  However, it is significantly
smaller than both the uncertainty from the fine structure constant and the
experimental uncertainty for $a_e$, see the discussion in
Ref.~\cite{Aoyama:2012wj}. 


\section*{Acknowledgements}

This work was supported by the DFG through the SFB/TR~9 ``Computational
Particle Physics''.  
P.M was supported in part by the EU Network LHCPHENOnet
PITN-GA-2010-264564 and HIGGSTOOLS PITN-GA-2012-316704.
We thank Hans K\"uhn for many discussions and for
initiating this project. We also thank Thomas Teubner for
carefully reading the manuscript and for many useful comments.
Furthermore, we would like to thank Thomas Teubner
and Daisuke Nomura for providing a {\tt FORTRAN} code for their compilation of
$R(s)$.  The Feynman diagrams were drawn with {\tt
  JaxoDraw}~\cite{Vermaseren:1994je,Binosi:2008ig}.


\section*{Note added:} 

During the refereeing process the paper~\cite{Colangelo:2014qya}
appeared on the arXiv. In that paper the NLO hadronic light-by-light
contribution, which is of the same perturbative order as the
corrections considered in our paper, has been estimated to
$a_\mu^{\rm lbl-had,NLO}=0.3 \pm 0.2 \times 10^{-10}$.



\end{document}